\documentstyle[12pt]{article}
\input psfig.sty
\newcommand{\ds}{\displaystyle}

\begin{document}

\title{Schr\"{o}dinger Equation with the\\ 
Potential $V(r)=Ar^{-4}+Br^{-3}+Cr^{-2}+Dr^{-1}$}

\author{Shi-Hai Dong\thanks{Electronic address: DONGSH@BEPC4. IHEP. AC. CN}\\
{\scriptsize Institute of High Energy Physics, P. O. Box 918(4), 
Beijing 100039, People's Republic of China}\\
\\
Zhong-Qi Ma\\
{\scriptsize China Center for Advanced Science and Technology
(World Laboratory), P. O. Box 8730, Beijing 100080}\\
{\scriptsize  and Institute of High Energy Physics, P. O. Box 918(4), 
Beijing 100039, People's Republic of China}}

\date{}

\maketitle

\begin{abstract}
Making use of an ${\it ansatz}$ for the eigenfunction, 
we obtain exact closed form solutions to 
the Schr\"{o}dinger equation with the inverse-power 
potential, $V(r)=Ar^{-4}+Br^{-3}+Cr^{-2}+Dr^{-1}$ 
both in three dimensions and in two dimensions, 
where the parameters of the potential $A, B, C, D$ 
satisfy some constraints. 



\end{abstract}

\newpage

\begin{center}
{\large 1. Introduction}\\
\end {center}

The exact solutions to the fundamental dynamical equations
play an important  role in the different fields of physics. 
As far as the  Schr\"{o}dinger equation concerned, 
the exact solutions are possible only for the several
potentials and some approximation methods
are frequently used to arrive at the solutions. 
The problem of the inverse-power potential, $1/r^n$, 
has been widely carried out on the different fields of
classic mechanics as well as on the
quantum mechanics. For instance, 
the interatomic interaction potential
in molecular physics [1-2], the 
inverse-power potentials $V(r)=-Z^2\alpha/r^4$ [3]
(interaction between an ion and a 
neutral atom) and $V(r)=-d_{1}d_{2}/r^3$ [4]
(interaction between a dipole $d_{1}$ 
and another dipole $d_{2}$) are often 
applied to explain the interaction 
between one matter and another one. The interaction in
one-electron atoms, muonic and 
hadronic and Rydberg atoms also requires considering the 
inverse-power potentials [5]. Indeed, the interaction
potentials mentioned above are only
special cases of the inverse-power potential
when some parameters of the potential vanishes. 

The reason we write this paper is as follows. 
On the one hand, 
\"{O}zcelik and Simsek discussed this
potential in the three-dimensional spaces [6]. 
They obtained the eigenvalues and
eigenfunctions for the arbitrary node. 
Simultaneously, the corresponding
constraints on the parameters of the 
potential were obtained. Unfortunately, 
they did not find that it is impossible 
to discuss the higher order excited
state except for the ground state. 
In the later discussion, 
we will draw this conclusion and find some essential
mistakes occurred in their calculations even for the ground state. 
We recalculate the
solutions to the Schr\"{o}dinger equation 
with this potential in three
dimensions following their idea and correct their mistakes. 
On the other hand, 
with the advent of growth
technique for the realization of
the semiconductor quantum wells, the
quantum mechanics of low-dimensional
systems has become a major research field. 
Almost all of the computational
techniques developed for the three-dimensional
problems have already been
extended to lower dimensions. 
Therefore, we generalize this method to the
two-dimensional Schr\"{o}dinger 
equation because of the wide interest in
lower-dimensional fields theory. 
Besides, we has succeeded
in dealing with the 
Sch\"{o}dinger equation with the anharmonic potentials, 
such as singular potential both in two dimensions 
and in three dimensions[9, 10], 
the sextic potential [11], 
the octic potential [12] and the Mie-type potential [13] 
by this method. 
We now attempt to 
study the Sch\"{o}dinger equation with 
the inverse-power potential 
by the same way both 
in three dimensions and in two dimensions. 

This paper is organized as follows. 
In section 2, we study the three-dimensional 
Schr\"{o}dinger equation with 
this potential using an ${\it ansatz}$ for the eigenfunctions. 
The study of the two-dimensional 
Schr\"{o}dinger equation with this potential will be
discussed in section 3. 
The figures for the unnormalized radial functions 
are plotted in the last section.

\begin{center}
{\large 2. Solutions in three dimensions }\\
\end{center}

Throughout this paper the natural unit $\hbar=1$ 
and $\mu=1/2$ are employed. Consider the 
Schr\"{o}dinger equation 
$$-\nabla^{2} \psi +V(r) \psi =E \psi, \eqno (1) $$ 

\noindent
where here and hereafter the potential
$$V(r)=Ar^{-4}+Br^{-3}+Cr^{-2}+Dr^{-1}, ~~~A>0, ~~~~D<0. \eqno(2)$$

\noindent
Let
$$\psi(r, \theta, \varphi)=r^{-1} R_{\ell}(r) Y_{\ell m}(\theta, \varphi), 
\eqno (3) $$  

\noindent
where $\ell$ and $E$ denote the angular momentum and the 
energy, respectively, and the radial wave function 
$R_{\ell}(r)$ satisfies
$$\displaystyle {d^{2} R_{\ell}(r) \over dr^{2} }
+\left [E-V(r)-\displaystyle {\ell(\ell+1) \over r^{2}} \right] 
R_{\ell}(r)=0. \eqno (4) $$

\noindent
\"{O}zcelik and Simsek [6] make an ${\it ansatz}$ 
for the ground state
$$R_{\ell}^{0}(r)=\exp [g(r)], \eqno(5)$$
 
\noindent
where
$$g(r)=\frac{a}{r}+b r+c \ln r, ~~~a<0, ~~~b<0. \eqno(6)$$

\noindent
After calculating, one can obtain the following equation 
$$\displaystyle {d^{2} R_{\ell}^{0}(r) \over dr^{2}}
-\left [\displaystyle{d^{2} g(r) \over dr^{2}}
+\left(\displaystyle {dg(r) \over dr}\right)^2\right]R_{\ell}^{0}(r)=0. 
\eqno(7)$$

\noindent
Compare Eq. (7) with Eq. (4) and obtain 
the following sets of equations
$$a^2=A, ~~~~~~b^2=-E, \eqno(8a)$$
$$2bc=D, ~~~~~2a(1-c)=B, \eqno(8b)$$
$$C+\ell(\ell+1)-\frac{1}{4}=c^2-2ba-c. \eqno(8c)$$

\noindent
It is not difficult to obtain the value
of the parameter $a$
from Eq. (8a) written as $a=\pm \sqrt{A}$. 
In order to retain the well-behaved solution
at $r \rightarrow 0$ and at $r \rightarrow \infty$, they
choose negative sign in $a$, i. e. $a=-\sqrt{A}$. 
According to this choice, 
they arrive at a constraint on the parameters of
the potential from Eq. (8c) written as
$$C=\ds{\frac{B^2}{4A}}+\displaystyle{\frac{B}{2\sqrt{A}}}
+\ds{\frac{2AD}{B+2\sqrt{A}}}-\ell(\ell+1). \eqno(9)$$

\noindent
Then the energy is read as
$$E_{0}^{\pm}=-\frac{1}{16A}\left\{C+\ell(\ell+1) \pm
\sqrt{ [C+\ell(\ell+1)]^2-2BD}\right\}^2. \eqno(10)$$

\noindent
It is readily to find that Eq. (10) is a wrong result. 
From Eqs. (6) and (8b), as we know, 
since the parameter $b$ is negative, 
when we calculate the energy $E$
from Eq. (8a), we only take 
the $b$ as a negative value, so that Eq. (10)
only takes the negative sign. 
Actually, it is not difficult 
to obtain the corresponding values of the
parameters for the $g(r)$ from Eq. (8), i. e. 
$$c=\displaystyle{\frac{B+2\sqrt{A}}{2\sqrt{A}}}, 
~~~b=\displaystyle{\frac{D\sqrt{A}}{B+2\sqrt{A}}}. \eqno(11)$$

\noindent
The eigenvalue $E$, however, 
will be simply expressed as from Eq. (8a)
$$E=-\ds{\frac{AD^2}{B^2+4A+4B\sqrt{A}}}. \eqno(12)$$

\noindent
The corresponding eigenfunction Eq. (5) can now be read as
$$R_{\ell}^{0}=N_{0}r^{c}
\exp\left [\frac{1}{r}a+b r\right], \eqno(13)$$

\noindent
where $N_{0}$ is the normalized constant
and here and hereafter the parameters 
$a$, $b$ and $c$
are given above. 

After their discussing the ground state, 
\"{O}zcelik and Simsek continue to
study the first excited state. 
They make the ${\it ansatz}$ for the
first excited state, 
$$R_{\ell}^{1}(r)=f(r)\exp [g(r)], \eqno(14)$$
 
\noindent
where $g(r)$ is the same as Eq. (6) and $f(r)=r-\alpha_{1}$, where
$\alpha_{1}$ is a constant. 
For short, it is readily to find from 
Eq. (14) that the radial wave function
$R_{\ell}^{1}(r)$ satisfies the following equation
$$R_{\ell}^{1}(r)''-\left [g(r)''+(g(r)')^2
+\left(\frac{f(r)''+2g(r)'f(r)'}{f(r)}\right)\right]
R_{\ell}^{1}(r)=0, \eqno(15)$$

\noindent
where the prime denotes the derivative of
the radial wave function with respect to the variable $r$. 

\noindent
Compare Eq. (15) with Eq. (4) and obtain
the following sets of equations
$$-2b-2bc+D+b^2\alpha_{1}+e\alpha_{1}=0, \eqno(16a)$$
$$-b^2-E=0, ~~~~~a^2\alpha_{1}-A\alpha_{1}=0, \eqno(16b)$$
$$-a^2+A+2a\alpha_{1}-B\alpha_{1}-2ac\alpha_{1}=0, \eqno(16c)$$
$$2ab-c-c^2+C+\ell(\ell+1)+2bc\alpha_{1}-D\alpha_{1}=0, \eqno(16d)$$
$$B+2ac-2ab\alpha_{1}-c\alpha_{1}+c^2\alpha_{1}-C\alpha_{1}-\ell\alpha_{1}-
\ell^2\alpha_{1}=0, \eqno(16e)$$

\noindent
it is not hard to obtain the
following sets of equations from Eqs. (16a-16c)
$$E=-b^2, ~~a^2=A, ~~c=\frac{B+2\sqrt{A}}{2\sqrt{A}}, \eqno(17a)$$
$$~~b=\frac{D\sqrt{A}}{B+4\sqrt{A}}, \eqno(17b)$$

\noindent
where the constant $\alpha_{1}\not=0$ and it is determined 
by Eqs. (16d) and (16e). Furthermore, 
it is evident to find that Eq. (17b) does not 
coincide with Eq. (11) with respect to the
same parameter $b$, which will lead to the their wrong calculation for the first
excited state. In fact, 
they obtained two different relations during their
calculation through the compared equation, 
i. e. $D=2bc$ (see Eq. (9) in [6])
and $D=2b(c+1)$ (see Eq. (16) in [6]). 
The parameter $D$ does not
exist if the parameter $b$ is not equal to zero. 
It is another main mistaken that arises their wrong result,  
that's to say, it is impossible to discuss the first excited
state for the Schr\"{o}dinger equation by this method. 
We only discuss the ground state
by this simpler ${\it ansatz}$ method as mentioned above. 

As a matter of fact, the normalized 
constants $N_{0}$ can be
calculated in principle from the normalized relation
$$\int_{0}^{\infty}|R_{\ell}^{0}|^2dr=1. \eqno(18)$$

\noindent
In the course of calculation, making use of
the standard integral [14](Re$\lambda_{1}>0$, 
Re$\lambda_{2}>0$ and Re$\nu>0$)
$$\int_{0}^{\infty}r^{\nu -1}\exp [-(\lambda_{1} r+\lambda_{2} r^{-1})]dr
=2\left(\frac{\lambda_{2}}{\lambda_{1}}\right)^{\nu/2}K_{\nu}
(2\sqrt{\lambda_{1}\lambda_{2}}), \eqno(19)$$

\noindent
which implies
$$N_{0}=\left [\ds{\frac{1}{2(\frac{a}{b})^{\frac{2c+1}{2}}
K_{2c+1}(4\sqrt{ab})}}\right], \eqno(20)$$

\noindent
where the values of the parameters $b, c$ and $a$ are given by Eq. (11) and
$-\sqrt{A}$, respectively. 
The figure 1 for the unnormalized radial eigenfunction in three dimensions 
is plotted in the last section. 

\begin{center}
{\large 3. Solutions in tow dimensions }\\
\end{center}

We now generalize this method to the two-dimensional 
Schr\"{o}dinger equation. Consider Schr\"{o}dinger 
equation with a potential $V(r)$
that depends only on the distance $r$ from the origin
$$H\psi =\left(
\displaystyle {1 \over r} \displaystyle {\partial \over \partial r} 
r \displaystyle {\partial \over \partial r} + 
\displaystyle {1 \over r^{2}} \displaystyle {\partial^{2} \over 
\partial \varphi^{2} } \right)\psi +V(r) \psi =E \psi. \eqno(21)$$ 

\noindent
Let
$$\psi(r, \varphi)=r^{-1/2} R_{m}(r) e^{ \pm im \varphi}, 
~~~~~m=0, 1, 2, \ldots, \eqno (22) $$  

\noindent
where the radial wave function 
$R_{m}(r)$ satisfies the following radial equation
$$\displaystyle {d^{2} R_{m}(r) \over dr^{2} }
+\left [E-V(r)-\displaystyle {m^{2}-1/4 \over r^{2}} \right] R_{m}(r)
=0, \eqno (23) $$

\noindent
where $m$ and $E$ denote the 
angular momentum and energy, respectively. 
For the solution of Eq. (23), 
we make an ${\it ansatz}$ [6-13] 
for the ground state
$$R_{m}^{0}(r)=\exp [g_{m}(r)], \eqno(24)$$
 
\noindent
where
$$g_{m}(r)=\frac{a_{1}}{r}+b_{1} r+c_{1} \ln r. \eqno(25)$$

\noindent
After calculating, we arrive at the following equation
$$\displaystyle {d^{2} R_{m}^{0}(r) \over dr^{2}}
-\left [\displaystyle{d^{2} g_{m}(r) \over dr^{2}}
+\left(\displaystyle {dg_{m}(r) \over dr}\right)^2\right]R_{m}^{0}(r)=0. 
\eqno(26)$$

\noindent
Compare Eq. (26) with Eq. (23) and obtain the following
sets of equations
$$a_{1}^2=A, ~~~~~~b_{1}^2=-E, \eqno(27a)$$
$$2b_{1}c_{1}=D, ~~~~~2a_{1}(1-c_{1})=B, \eqno(27b)$$
$$C+m^2-\frac{1}{4}=c_{1}^2-2b_{1}a_{1}-c_{1}. \eqno(27c)$$

\noindent
It is not difficult to obtain the values
of the parameters $a_{1}$
from Eq. (27a) written as $a_{1}=\pm \sqrt{A}$. 
Likely, in order to retain the well-behaved solution
at $r \rightarrow 0$ and at $r \rightarrow \infty$, we
choose negative sign in $a_{1}$, i. e. $a_{1}=-\sqrt{A}$. 
According to this choice, 
Eq. (27b) will give the other parameter values as
$$c_{1}=\displaystyle{\frac{B+2\sqrt{A}}{2\sqrt{A}}}, 
~~~b_{1}=\displaystyle{\frac{D\sqrt{A}}{B+2\sqrt{A}}}. \eqno(28)$$

\noindent
Besides, it is readily to obtain from Eq. (27c) that
$$C=\ds{\frac{B^2}{4A}}+\displaystyle{\frac{B}{2\sqrt{A}}}
+\ds{\frac{2AD}{B+2\sqrt{A}}}-(m^2-1/4), \eqno(29)$$

\noindent
which is the constraint on 
the parameters for the two-dimensional 
Schr\"{o}dinger equation with the inverse-power potential. 

\noindent
The eigenvalue $E$, however, will be given by Eq. (27a) as
$$E=-\ds{\frac{AD^2}{B^2+4A+4B\sqrt{A}}}. \eqno(30)$$

\noindent
The corresponding eigenfunction Eq. (24) can now be read as
$$R_{m}^{0}=Nr^{c_{1}}
\exp\left [\frac{1}{r}a_{1}+b_{1} r\right], \eqno(31)$$

Similarly, the normalized 
constants $N$ can be
calculated in principle from the normalized relation
$$\int_{0}^{\infty}|R_{m}^{0}|^2dr=1. \eqno(32)$$

\noindent
According to Eq. (19), we can obtain
$$N=\left [\ds{\frac{1}{2(\frac{a_{1}}{b_{1}})^{\frac{2c_{1}+1}{2}}
K_{2c_{1}+1}(4\sqrt{a_{1}b_{1}})}}\right], \eqno(33)$$

\noindent
where the values of the parameters 
$a_{1}, b_{1}$ and $c_{1}$ are given above. 
The figure 2 for the unnormalized radial eigenfunction in two dimensions 
is plotted in the last section. 

Considering the values of the
parameters of the potential, we fix them as follows. 
The values of parameters $A, C, D$ are first fixed,
for example $A=4. 0, C=2. 0$ and $D=-2. 0$,
the value of the parameter $B$
is given by Eq. (10) and Eq. (29) for the
cases both in three dimensions and 
two dimensions for $\ell=0$ and $m=0$, respectively. 
By this way,
the parameter $B$ turns out to $B=5. 87$ 
in three dimensions and  $B=5. 65$ in two dimensions, 
respectively.  
The ground state energy corresponding to these
values are obtained as $E=-0. 164$ for the case
in three dimensions and $E=-0. 172$ for the case in two dimensions. 
Actually, when we study the properties
of the ground state, as we know, the unnormalized
radial wave functions do not
affect the main features of the wave functions. 
We have plotted the unnormalized
radial wave functions in figures 1 and 2
for the cases both in three 
dimensions and in two dimensions, respectively. 
With respect to figures 1 and 2, 
it is easy to find that they are similar to 
each other, which 
stems from the same values of the  angular momentum 
$\ell=0$ and $m=0$. They will be 
different if we take the different 
values of the angular momentum 
in the course of calculations. 

In conclusion, we obtain the exact 
analytic solutions to the Schr\"{o}dinger
equation with the inverse-power potential 
$V(r)=Ar^{-4}+Br^{-3}+Cr^{-2}+Dr^{-1}$ using
a simpler ${\it ansatz}$ for the eigenfunction 
both in three dimensions and in two dimensions, 
and simultaneously 
the constrains on the parameters of the potential
are arrived at from the 
compared equations. Finally, we remark that this simple and 
intuitive method can be generalized 
to other potential. The study of the
Schr\"{o}dinger equation with the 
asymmetric potential is in progress.

\vspace{10mm}
{\bf Acknowledgments}. This work was supported by the National
Natural Science Foundation of China and Grant No. LWTZ-1298 from
the Chinese Academy of Sciences.

\newpage

\vskip 1cm
\begin{figure}
\begin{center}
\mbox{\psfig
{figure=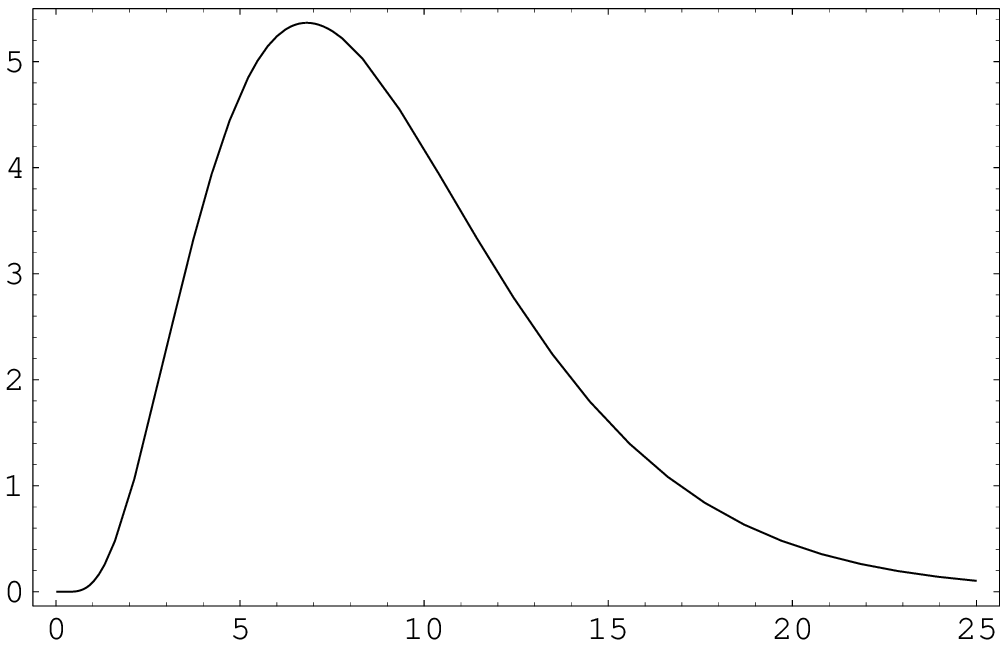,height=5cm,width=8cm}}
\end{center}
\caption{The ground state wave functions in three dimensions as
a function of $r$ for the
potential (2) with the values $A=4. 0, B=5. 87, C=2. 0$
and $D=-2. 0$.
The $y$-axis denotes the values of wave
functions and the $x$-axis denotes the
variable $r$. } 
\end{figure}

\vskip 1cm
\begin{figure}
\begin{center}
\mbox{\psfig
{figure=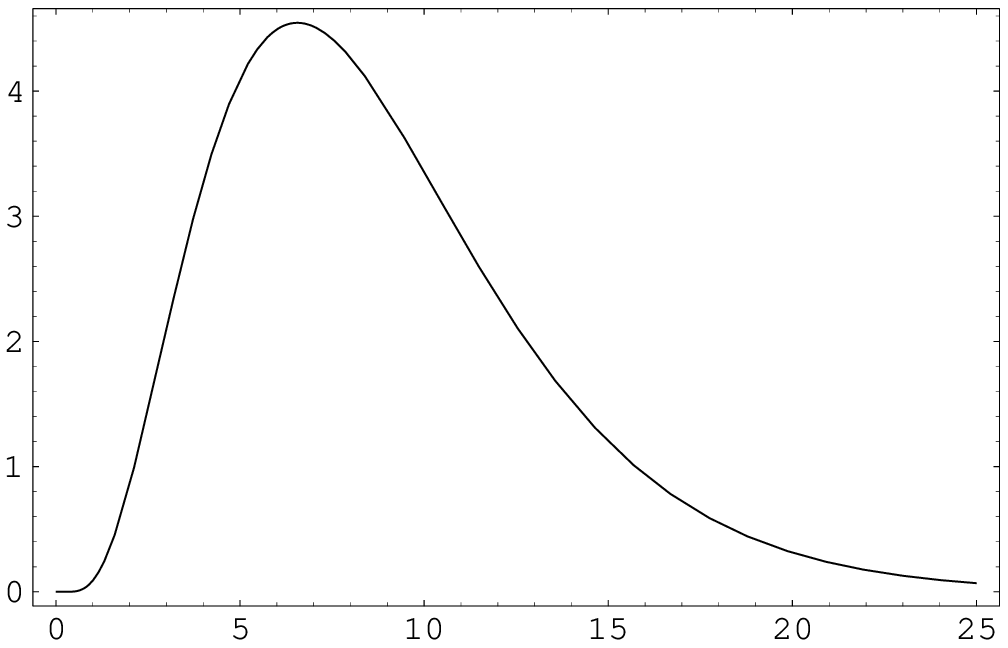,height=5cm,width=8cm}}
\end{center}
\caption{The ground state wave functions in two dimensions as
a function of $r$ for the
potential (2) with the values
$A=4. 0, B=5. 65, C=2. 0$
and $D=-2. 0$. The $y$-axis denotes the values of wave
functions and the $x$-axis denotes the
variable $r$. } 
\end{figure}

\end{document}